\documentstyle[11pt, epsfig]{article}
\newcommand{\blankline}{\vskip .3cm}
\newcommand\beq{\begin{equation}}
\newcommand\eeq{\end{equation}}
\begin{document} 
\centerline{\Large\bf Dual formulation of spin network evolution}
\vspace{0.5in}
\rm
\centerline{Fotini Markopoulou}
\blankline
\centerline{\it  Center for Gravitational Physics and Geometry}
\centerline{\it Department of Physics}
 \centerline {\it The Pennsylvania State University}
\centerline{\it University Park, PA, USA 16802  $ {}^*$ }
\vfill
\centerline{April 5, 1997}
\vfill 
\centerline{\bf Abstract}
\blankline
{\small
We illustrate the relationship between spin networks and their dual 
representation by labelled triangulations of space in 2+1 and 3+1 dimensions. 
We apply this to the recent proposal for causal evolution of spin
networks. The result is labelled spatial triangulations evolving with
transition amplitudes given by labelled spacetime simplices. 
The formalism is very similar to simplicial gravity, however, the
triangulations represent combinatorics and not an approximation to the
spatial manifold. 
The distinction between future and past nodes which can be ordered in causal sets also exists here. Spacelike and timelike slices can be defined and the foliation is allowed to vary.
 We clarify the choice of the two rules in the causal spin network
evolution, and the assumption of trivalent spin networks for 2+1
spacetime dimensions and four-valent for 3+1. As a direct application,
the problem of the exponential growth of the causal model is remedied.
The result is a clear and more rigid graphical understanding of
evolution of combinatorial spin networks, on which further work  can
be based. }

\vskip 1cm
${}^{*}$ {\small Permanent address: 
Theoretical Physics Group, Blackett Laboratory, 
Imperial College of Science, Technology and Medicine, London SW7 2BZ,
U.K.\\
\indent email address:  fotini@phys.psu.edu}
\eject

\section{Introduction}

A model for the causal evolution of spin networks in quantum gravity
has recently been proposed by Markopoulou and Smolin \cite{MaSm}. This
proposal advocates exploiting the discreteness of space suggested by
the appearance of spin networks in canonical gravity (where their
diffeomorphism classes have been shown to provide a basis for the
kinematical state space
\cite{{abhay1},{tedlee},{lp1},{lp2},{ls-review},{abhay-book},{carlo-review}})
in order to 
address the issue of time evolution. The model lies between the loop
formulation of canonical gravity and the causal set
approach\footnote{For the present purposes a causal set is a set of
points which has the causal properties that may be assigned to sets of
points in Minkowskian spacetime: to each pair either one is to the
future of the other, or they are causally unrelated. } \cite{sorkin}
since it assigns quantum amplitudes to special classes of causal sets
which consist of spin networks representing quantum states of the
gravitational field joined together by labelled null edges.  
 The dynamics is specified by a choice of functions of the labellings
of $d+1$ dimensional simplices which represent elementary future light
cones of events in these discrete spacetimes.  

More specifically, given an initial spin network $\Gamma_1$, there are
two rules for the amplitudes of the transition from $\Gamma_1$ to the
next spin network $\Gamma_2$. These rules do two things. First, they
allow all possible moves from the spin network at one time instant to
the later one, which is the creation of new edges (rule 1) or the
recoupling of existing ones (rule 2), thus making use of the
combinatorial nature of the spin networks. Second, the rules attempt
to set up a causal structure in the resulting ``spacetime network'' --
the combination of the spin networks that make up the spatial slices
and the null edges (the lightcones) that connect them.  

Given the two spin networks $\Gamma_1$ and $\Gamma_2$, the amplitude
for the transition from $\Gamma_1$ to $\Gamma_2$, ${\cal
A}_{\Gamma_1\rightarrow\Gamma_2}$ can be constructed by applying these
two rules. Let ${\cal G}$ be the collection of all causal spacetime
networks consistently built by the alternation of the two rules which
has $\Gamma_i$ as the zeroth spin network and  $\Gamma_f$ as the final
one, and $L_{\cal G}$ the number of such spin networks $\Gamma_I$ in
${\cal G}$. Then the transition amplitude from $\Gamma_i$ to
$\Gamma_f$ is 
\beq
{\cal A}_{\Gamma_i\rightarrow\Gamma_f}=\sum_{\cal G}
\prod^{L_{\cal G}-1}_{I=0}{\cal A}_{\Gamma_I\rightarrow\Gamma_{I+1}}
\eeq
where the sum is over all allowed labellings and the amplitude is
defined alternatively in terms of the two rules (see \cite{MaSm} for
more details).  

This model was not derived from the classical theory, rather it is an
attempt at constructing a transition amplitude between spin network
states that is consistent with some discrete microscopic form of
causality. Further,  it displayed certain features of theories with
critical behaviour, 
indicating that a renormalization group approach may be applicable  
(as it has been discussed in \cite{ls}).   

Constructing a model of causal spin network evolution turned out to be
a mixture of combinatorial spin networks, causal sets and critical
behaviour models. Presumably, progress on all of these features should
be made before satisfactory resuls come out. In this paper, we take up
the first aspect, the combinatorial structure of the spin networks as
it was originally suggested by Penrose \cite{penrose}
(which may be interpreted 
as implying a fundamentally discrete space (and spacetime)).
Indeed, the present application is part of work in progress where the
appearance of spin networks in canonical gravity is regarded as
evidence for discrete structure of space(time) and is 
viewed from  the perspective of category theory which we believe is
the relevant mathematical formalism.  An immediate result is the
duality between spin networks and triangulations, which we overview
here and then apply it to the  
causal scheme of spin networks evolution, thus clarifying the 
evolution rules. We also describe how this viewpoint allows a simple
remedy for the exponential growth in the causal model.

\section{Spin networks and triangulations}

In the causal evolution of spin networks \cite{MaSm} a spin network is
regarded as a combinatorial labelled graph with nodes and edges
labelled according to the rules satisfied by spin networks
\cite{rosm, penrose}. The edges are labelled by representations of
SU(2) and the 
nodes by intertwiners, which are distinct ways of extracting the
identity representation from the products of representations of the
incident edges. The spin networks are defined only by their
combinatorics and no embedding in a spatial manifold is assumed.  
 
A more rigid and reliable formalism,  very useful when one needs to
rely heavily on the combinatorics of spin networks, can be obtained if
they are treated as the 1-skeletons of $n$-simplices triangulating the
$n$-dimensional spatial manifold.  This correspondence may be found,
for example, in topological quantum field theory literature
\cite{craneetal}, and in observations on the relationship between
spin networks and simplicial gravity \cite{HaPe}. A basic difference
is that we treat triangulations exactly like their dual spin networks,
namely as combinatorial constructions. This is in contrast to their
role in topological quantum field theory and simplicial gravity where
the triangulation is an approximation to the spatial manifold. 
Further, this correspondence has not been explored in spin network
canonical gravity since the emphasis  has been on spin networks
embedded in the three-dimensional space (spin network states are the
basis for cylindrical functions, the elements of the Hilbert space of
kinematical states) 
rather than regarding the appearance of spin networks as evidence for
discrete, and possibly combinatoric, structure. 
  
The relevant spatial triangulations and spacetime simplices are different 
 in 2+1 an 3+1 dimensions. We describe the two sets of spatial
triangulations in the next two subsections and in section 3 we
describe their spacetime evolution.

\subsection{2-dimensional space}

\begin{figure}
\centerline{\mbox{\epsfig{file=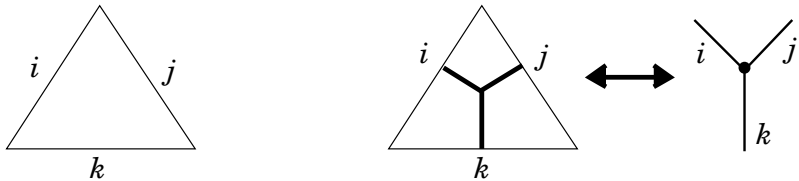}}}
\caption{ The 2-simplex triangulating 2-dimensional space and its dual
trivalent spin network. The edges of the triangle are labelled by
spins and its face by an intertwiner (omitted in an SU(2) spin
network). } 
\label{triangle}
\end{figure}

In 2+1 spacetime dimensions, the simplex triangulating the
2-dimensional space is a triangle.  One triangle corresponds to one
node of the spin network, and its sides are labelled by the same
colors as this dual spin network (figure \ref{triangle}). This implies
that the spin networks employed in the description of 2-dimensional
space are restricted to be  {\it trivalent}.\footnote{
A $n$-valent node in 2-dimensional space would be dual to a
$n$-gon. This can be divided into triangles by the same method that an
$n$-valent node can be split into trivalent ones by introducing
virtual edges.}  In general, the faces of
the triangles are labelled by intertwiners. In the case of SU(2) spin
networks they are trivial and hence the label does not appear.  

\begin{figure}
\centerline{\mbox{\epsfig{file=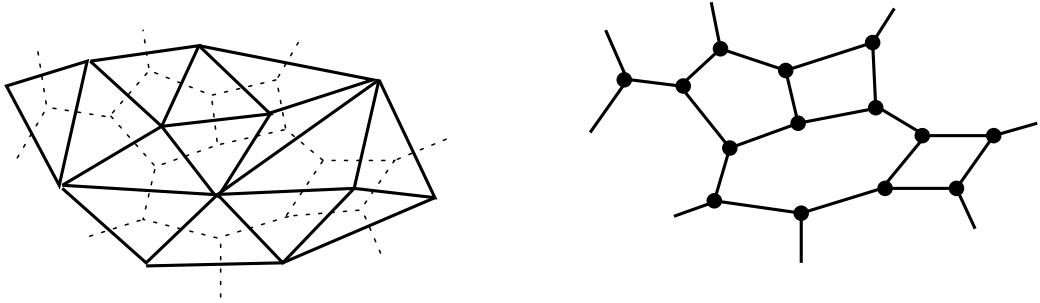}}}
\caption{A 2-dimensional spatial manifold can be described by a
(labelled) triangulation (left), or its dual trivalent spin network
(right). } 
\label{triangulation}
\end{figure}

\begin{figure}
\centerline{\mbox{\epsfig{file=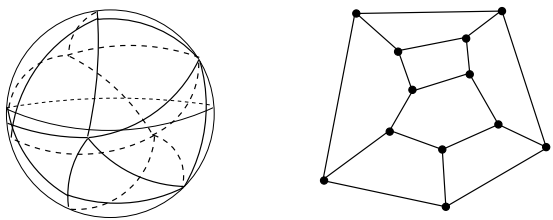}}}
\caption{An example of a triangulated closed surface. It gives rise to
a closed spin network.} 
\label{closed}
\end{figure}

This provides a triangulation of 2-dimensional space that corresponds
to its kinematical description  by a trivalent spin network (figure
\ref{triangulation}). Note that a closed spin network corresponds to
closed space (figure \ref{closed}).

\subsection{3-dimensional space}

\begin{figure}
\centerline{\mbox{\epsfig{file=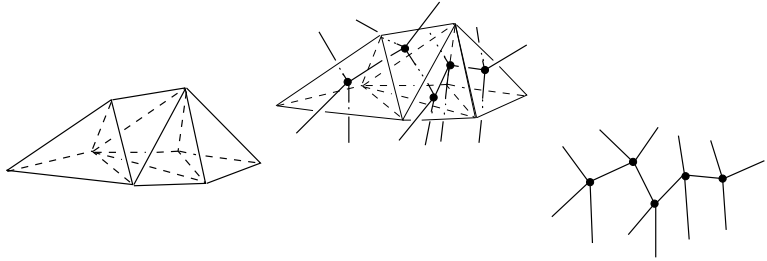}}}
\caption{Part of a four-valent spin network, five nodes, corresponds
to five teterahedra of a triangulation of 3-dimensional space. } 
\label{T}
\end{figure}

\begin{figure}
\centerline{\mbox{\epsfig{file=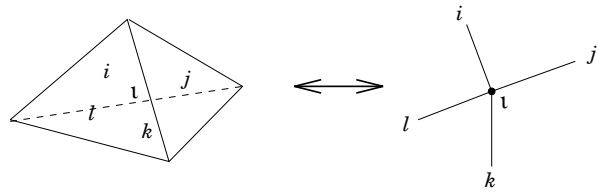}}}
\caption{The labelling of a tetrahedron used in a triangulation of
3-dimensional space. Its faces are labelled by spins and the
tetrahedron itself by an intertwiner. } 
\label{Tlabels}
\end{figure}

The 3-dimensional space is triangulated by tetrahedra. Again the  dual
spin network is the 1-skeleton of the tetrahedra, so that there is one
node for each tetrahedron  and one spin network edge puncturing each
of its faces (figure \ref{T}). This means that we now allow only {\it
four-valent} spin networks. The tetrahedra are labelled by
intertwiners and their faces by spins (figure \ref{Tlabels}).  

\vskip 0.5cm

It is most intriguing that one can now propose a straightforward
correspondence between spins puncturing faces of tetrahedra and the
area of those faces, or the intertwiners labelling the tetrahedra and
their volume, like the standard spin network results on area and
volume \cite{areavol}.  We will investigate this and the relationship to
simplicial gravity and Regge calculus it suggests in future work.  

Having now discussed the relevant triangulation of space, in the next
section we describe the corresponding picture of causal time
evolution.

\section{Spacetime evolution and causality}

We shall discuss this mainly in 2+1 dimensions, as the 3+1 case
involves 4-dimensional spacetime simplices which are rather hard to
visualize. At the level of the present discussion, the observations in
3+1 are completely analogous to those of 2+1, and thus none of the
important insights are lost with 
this simplification. 

\subsection{2+1 evolution}

In 2+1 spacetime dimensions, evolution will be a transition from a
triangulation of the 2-dimensional space  at a given time instant to a
later one. The way to do this is by placing spacetime tetrahedra,
whose faces are the triangles described in 2.1, on top of the initial
triangulation. (These are {\it spacetime} tetrahedra and unrelated to
the spatial ones dual to four-valent spin networks in 3+1 dimensions).

\begin{figure}
\centerline{\mbox{\epsfig{file=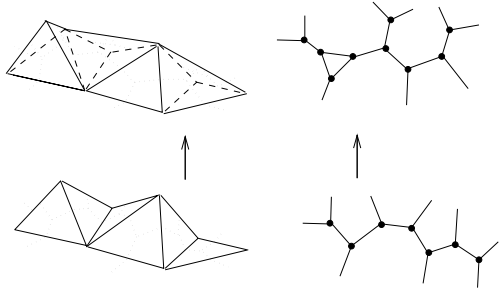}}}
\caption{Evolution of a spin network in 2+1 spacetime. }
\label{2+1evol}
\end{figure}

 The tetrahedra can then be regarded as a triangulation of
spacetime. An initial slice is the spin network drawn on the faces of
the initial spatial triangulation. A time step consists of tetrahedra
being placed on this triangulation, their bottom faces being the
triangles of the initial triangulation. On their top faces, the new
triangles, we can again draw the dual spin network. This is the new
spatial slice (figure \ref{2+1evol}). (Whether there is a change of
all the nodes by such a placing of a tetrahedron in each time step is
discussed in section 4.) 

Note that we define a time step to last the entire ``height'' of a
spacetime simplex. That is, in this model spatial slices do not cut
through the simplices. One can also construct discrete theories where
this happens, and may be necessary to consider such cases. However,
this is not something we shall 
investigate here (see, for example, the approach of J.~Zapata in
\cite{jose}).

In the spin network model of \cite{MaSm}, causality was built in  by
introducing a distinction between future nodes, produced by
intersections of lightcones of past nodes, and those past nodes. Here
this translates to the distinction between future and past {\it faces}
of the spacetime tetrahedron. As it is placed on the spatial
triangulation, the bottom faces correspond to  its past nodes and the
top to the future nodes. Then the labels on future faces and edges
depend only on the past faces and edges of the same tetrahedron. Thus,
there is a natural partial ordering of the faces and we can again
arrange them into causal sets.  

\begin{figure}
\centerline{\mbox{\epsfig{file=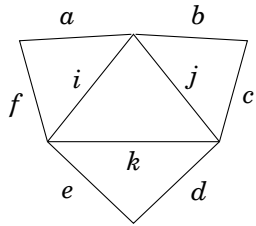}}}
\caption{Local evolution.}
\label{local}
\end{figure}

The model is {\it local} in the sense that if at time $t_1$ the four
triangles of figure \ref{local} are causally unconnected, i.e.\ an
observer in the middle triangle is not communicating with the three
adjacent ones, in the next step $t_2$ all four triangles in the region
bounded by spins $a,...,f$, but no more, will be in touch with each
other.  

Further, having seen what a spatial slice looks like and what is the
causal ordering of the nodes, we can now give a definition of a {\it
spacelike} slice:  a spacelike slice should not combine future and
past nodes of the same spacetime simplex, i.e.\ a spacelike slice
should not wrap around a spacetime simplex. In contrast, {\it
timelike} surfaces will be those that do contain both future and past
nodes of the same spacetime simplex. It is important to note that this
allows a varying ``foliation'' of spacetime, subject to the slices
that  follow the given initial surface being spacelike.  

\begin{figure}
\centerline{\mbox{\epsfig{file=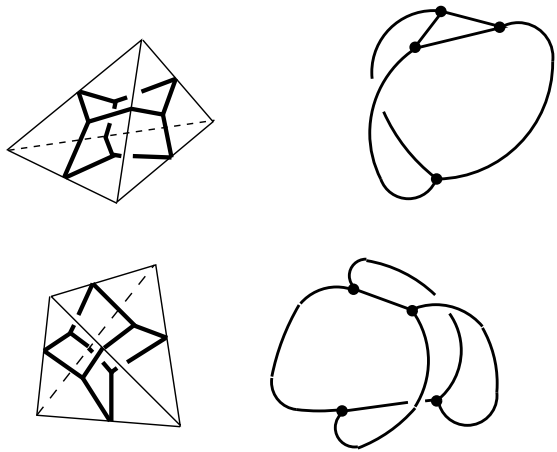}}}
\caption{Top: a spacetime tetrahedron with one face down and its dual
spin network amplitude. Bottom: We get the amplitude for recoupling of
edges from a tetrahedron standing on one edge. } 
\label{6j}
\end{figure}

\begin{figure}
\centerline{\mbox{\epsfig{file=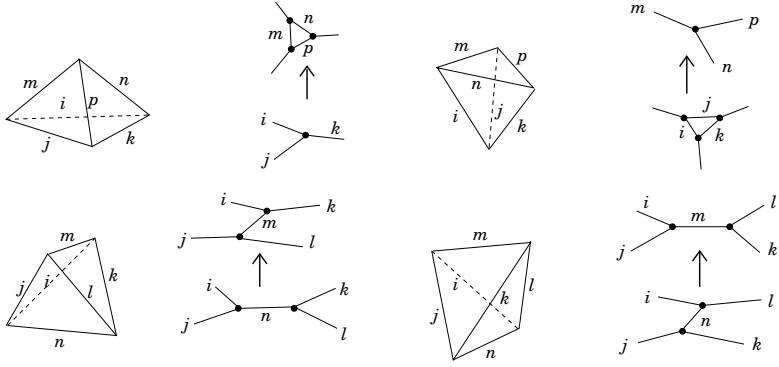}}}
\caption{The 1--3, 3--1, and the two 2--2 possibilities. The bottom
faces correspond to past nodes and the top to the future. } 
\label{2+1amps}
\end{figure}

Let us now write down conditions on the transition amplitudes from an initial to a final spin network that may be inferred from this formulation. 
  The spacetime tetrahedron is the amplitude for the possible
transitions between past and future nodes and can be evaluated using
an appropriate type of $6j$-symbol. 
It has 4 faces, hence 4 dual nodes, and the various ways it can be
placed on the past triangulation correspond to either 2 past nodes
going to 2 new ones, or 1 past node to 3 new and vice versa (fig.\
\ref{6j}).  The 2--2 case is a tetrahedron standing on one edge and
gives rise to recoupling of edges. There are two 2--2 possibilities,
fig.\ \ref{2+1amps}, which we take to have the same weight.  
A tetrahedron with one face down will be the process where one node
splits to a triangle of three new ones (1--3), while the inverted
version of this will be the opposite process (3--1) (fig.\
\ref{2+1amps}).  

It is clear that these are all the possibilities that can arise when
tetrahedral amplitudes are placed on a trivalent spin network. Note
that the result is the same as Rules 1 and 2 of the causal evolution
of spin networks. Rule 1 produced three new nodes/edges for each past
node, thus it is the 1--3 move. Rule 2 allowed for all possible
recouplings of the new spin network, hence it is the 2--2 case. We can
now see that the choice of those rules is justified, as they are
simply the full set of possibilities for evolution of a combinatoric
trivalent spin network. We can also see that the 3--1 case was omitted
in \cite{MaSm}. In section 4 we show that this omission is responsible
for the problematic exponential growth of the spin network in that
case.  

The important issue is to find which amplitudes should be assigned to
the spacetime tetrahedra to give us the new labels as functions of the
old ones, satisfying our requirements of local and causal
evolution. This remains unresolved. However, the construction itself
provides some guidance. First, we are using combinatorial spin
networks, hence we are looking for $6j$-type amplitudes, i.e\
combinatorial expressions for the new labels. Second, the spacetime
simplices are {\it directed} because we distinguish between their
future and past faces. Consequently, the amplitudes we are looking for
will have less symmetry than the Regge-Ponzano $6j$-symbols.  

Let us then name the 1--3 amplitude $J(mnp;ijk)$, $m,n,p$ being the
new labels and $i,j,k$ the old ones. We are then allowed to permute
each of the two sets $m,n,p$ and $i,j,k$ but not to mix
them. Similarly we can call the 3--1 amplitude $J'(mnp;ijk)$. In the
2--2 case (see fig.\ \ref{2+1amps}), we have recoupling that sends an
old edge $n$ into a new one $m$, keeping the outer labels $i,j,k,l$
fixed. Presumably, in this case we can only permute $i,j,k,l$, hence
let us name this amplitude $P(m;ijkl;n)$ and the other 2--2 case
$P'(m;ijkl;n)$. 

A {\it site} in a given spatial triangulation is where one spacetime
tetrahedron is placed and it may contain one, two, or three
triangles. If it only has one, then the amplitude $J'$ applies, if two
then it's either $P$ or $P'$, and if it has three triangles we need
$J$. Selecting the right amplitude is the same as selecting the site,
hence we do not need to count both. As a result, the amplitude to go
from a given spin network $\Gamma_0$ to the next one $\Gamma_1$ is
given by 
\beq
{\cal A}_{\Gamma_0\rightarrow\Gamma_1}=\prod_{\rm sites\ in\ the\
triangulation\  dual\  to\  {\Gamma_0}} A({\rm site}), 
\label{eq:a1}
\eeq
where $A({\rm site})$ can be $J,\ J',\ P,$ or $P',$ according to the
type of site. 

If, between an initial and a final spin network $\Gamma_i$ and $\Gamma_f$, 
or equivalently the boundaries of a given spacetime triangulation, a
foliation of spacelike slices contains a finite number $N$ of spin
networks, then the amplitude from $\Gamma_i$ to $\Gamma_f$ is 
\beq
{\cal A}_{\Gamma_i\rightarrow\Gamma_f}|_{\rm fixed\ triangulation}=
\prod^{I=N-1}_{I=0}{\cal A}_{\Gamma_I\rightarrow\Gamma_{I+1}}.
\eeq
And if we further allow the spacetime triangulation to vary then the
overall amplitude is given by 
\beq
{\cal A}_{\Gamma_i\rightarrow\Gamma_f}=
\sum_{\rm possible\ spacetime\ triangulations}
\prod^{I=N-1}_{I=0}{\cal A}_{\Gamma_I\rightarrow\Gamma_{I+1}}.
\label{eq:a3}
\eeq
Implementing causality now involves finding the correct choices of
these amplitudes, which at this stage is an unsolved problem. 

\subsection{3+1 evolution}

\begin{figure}
\centerline{\mbox{\epsfig{file=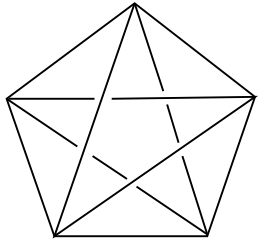}}}
\caption{The 4-simplex, the spacetime transition amplitude for
3-dimensional spatial triangulations.} 
\label{4simplex}
\end{figure}

\begin{figure}
\centerline{\mbox{\epsfig{file=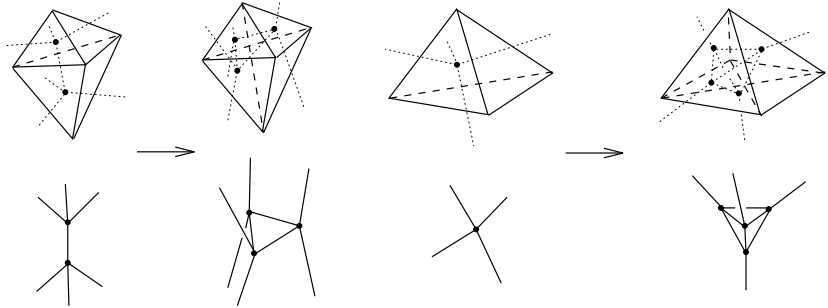}}}
\caption{The Pachner moves. The 4-simplex contains 5 tetrahedra hence
5 dual nodes. We show the cases of 2 past to 3 future nodes and 1 past
to 4 future nodes.  } 
\label{pachner}
\end{figure}

In 3+1 dimensions one repeats the above construction for  the
three-dimensional spatial manifold triangulated by tetrahedra
(four-valent spin networks). The spacetime is now triangulated by the
4-simplex of figure \ref{4simplex}. It is made up of  10 faces
labelled by spins and 5 tetrahedra labelled by intertwiners. Placing
4-simplices on top of the spatial tetrahedra again gives us a
distinction between past and future nodes. Since it has 5 tetrahedra,
and hence 5 dual nodes which can be separated to past and future ones,
different placements of the 4-simplex represent the transition
amplitudes 1--4 of 1 past node to 4 future, 2--3, two past nodes to 3
future, and vice versa (the issue of the appropriate $15j$ symbols
that respect causality also applies here). These are the Pachner
moves, shown in figure \ref{pachner}. 

Again, the use of 4-simplices and Pachner moves has the same effect as
applying Rules 1 and 2 to four-valent spin networks, as was proposed
in \cite{MaSm}. There, no justification for the choice of four-valent
spin networks for 3+1 dimensions was made, but this is now simply the
spin network dual to a triangulation of the 3-dimensional
space. (Again nodes of higher valence than four corresponding to
polyhedra can be broken down to fourvalent ones by reducing the
polyhedron to tetrahedra.) 

\vskip 0.5cm
Thus it becomes clear how the features introduced in \cite{MaSm} of a discrete spacetime described by a local, causal evolution of combinatorial spin networks relate to a triangulation of space and spacetime. 
In the next subsection we discuss further how formulating this
problem in terms of triangulations helps to clarify certain aspects of
causal evolution.   

\subsection{Space and spacetime}
 
The existing proposals for evolving spin networks often come under the
names of spacetime or 4-dimensional formalisms. An issue here is
whether one should extend spin networks to 4 dimensions and thus have
a ``4-dimensional spin network'' or restrict spin networks to
represent space only and obtain the 4-dimensional theory in some other
way. A further confusion arises because of the similarities between
3-dimensional space and 2+1 Euclidean gravity.  

The 4-dimensional models involving spin networks in the
literature---apart from the spacetime network of  Markopoulou and
Smolin being considered here---are those of Reisenberger and Rovelli
\cite{RR} and Baez \cite{foam} which feature a spacetime construction
using surfaces.\footnote{There is also the simplicial model for
euclidean general relativity of Reisenberger \cite{mike}, and in
topological quantum field theory the work in 4 dimensions of Crane
\cite{{craneetal}, {crane}}, Ooguri \cite{ooguri}, and others.  In
particular, Reisenberger has already pointed out how the 
duality between spin networks and triangulations clarifies a
relationship between the causal evolution rules and topological
quantum field theory \cite{mike2}.} 
They can be seen as branching surfaces connecting the initial and
final spin networks, or, conversely, spin networks can be regarded as
slicings of the labelled spacetime surfaces (``spin
foams''). Presumably, some of the features of these models are truly
essential and possibly common in the different models, while others
serve mostly as visual aids or in calculational strategies. The
duality of spin networks to triangulations can help a great deal in
elucidating the essential features.  
 
\begin{figure}
\centerline{\mbox{\epsfig{file=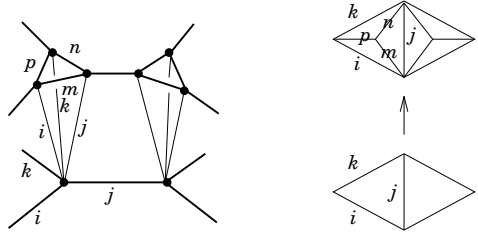}}}
\caption{The null edges duplicate labels of the initial spin
network. In the dual triangulation all labels are spacelike.}
\label{internet}
\end{figure}

We have seen that starting from an initial spin network $\Gamma_i$
dual to a triangulation, the next one is obtained by local changes of
the first, whose amplitudes can be represented by spacetime simplices
(this is illustrated in more detail in the next section). Now these
simplices clearly are amplitudes and not spin networks themselves. No
extra non-spatial edges are introduced by the spacetime simplex. That
is, all edges one can draw on a spacetime simplex belong to some
spatial slice.  
Therefore, the ``internet'' or null edges of \cite{MaSm} are not
really there.  
 They duplicate labels that were already in $\Gamma_i$.  To see this
let us consider figure \ref{internet}.  
Two nodes in the initial spin network $\Gamma_i$ go to six nodes in 
the next one, $\Gamma_f$. The null edges used  to 
write the transition amplitude ${\cal A}_{\Gamma_i\rightarrow\Gamma_f}$ are 
in fact labelled with the same spins as the spatial edges they come from 
(this is dictated by the causality requirement). Therefore, they duplicate 
existing labels. In the dual triangulation on the right of fig.\
\ref{internet}, all spin labels are  
spatial, i.e.\ they belong to the some spatial slice.

The null edges have one more important role, to keep track of
causality, namely determining the dependence of new labels on old. In
triangulations this is done automatically by identifying the top and
bottom faces (tetrahedra in 3+1) of a spacetime simplex to be the
future and past nodes. Also, timelike surfaces can still be defined,
as explained earlier in section 3.  

Keeping in mind that $(n-1)$-dimensional space is an $n$-valent spin
network dual to  spatial $(n-1)$-simplices, and that this spin network
evolves with amplitudes given by $n$-simplices (which are not
themselves associated with dual spin networks), spacetime built this
way may be thought of as made up of all evolution amplitudes in
equations (\ref{eq:a1})--(\ref{eq:a3}).  

\section{Controlling the growth of the spin network}

The evolution model described in \cite{MaSm} is the extreme case where
every node of the spin network is a spacetime event, that is, new
edges are created at every time step and for every node. One may
object that in this case the spin network evolves too fast.  However,
this can be easily controlled, as we will discuss in this section.
Again, for clarity,  we discuss the 2+1 case. There is no significant
difference in 3+1.  

There are, in fact, two ways to control growth. 
First, the problem can be fixed by simply noticing that the model did
not include the 3--1 case which we encountered in its translation into
triangulations, where three past nodes evolve to only one future node
(top right of figure \ref{2+1amps}). Evolution by placing a spacetime
simplex over each possible site at each time step (as in fig.\
\ref{2+1evol}) but including the 3--1 move means that the number of
nodes no longer increases exponentially.  

It may well be that this is all one needs to do. However, there is a
second intriguing option that arises here, and at present there is no
reason to prefer one of the two. This we call the non-maximal case,
and its treatment would be reminiscent of directed percolation. It
produces the effect of multifingered time. 

\begin{figure}
\centerline{\mbox{\epsfig{file=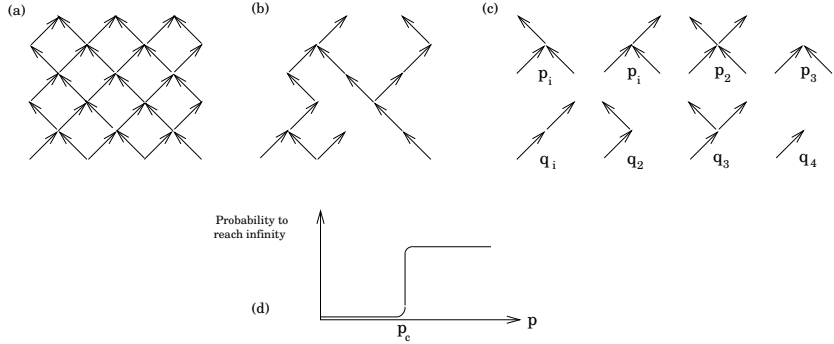}}}
\caption{ (a) Percolation lattice with all arrows on. (b) A history in
the percolation model. (c) A possible assignment of probabilities to
the various cases; $2p_1+p_2+p_3=1, \ \ \Sigma_i q_i=1$. (d)
Probability to reach infinity against the probabilities $p$ assigned
to the nodes.} 
\label{perc}
\end{figure}

Let us recall the example of the statistical model of directed
percolation, most easily seen in its 1+1 form \cite{perc}. The causal
links (null edges between nodes) are represented (in figure
\ref{perc}) as future pointing arrows. Each arrow can be on or off and
there can be a probability amplitude $p$ associated with the various
possibilities. One can understand this model as information flowing
from the bottom to the top, with $P$ being the probability that
information from the bottom succeeds in flowing through to the top.  
The critical behaviour of this model may be seen by graphing $P$
against the probability $p$ assigned to each node (essentially the
number of arrows that are on).
 In the extreme case where
all arrows are on, $P$ is of course 1. As the number of arrows on
decreases, $P$ remains close to 1, down to some critical value
$p_c$. A short while after $p_c$ is crossed, information flow fails to
make it to the top (fig.\ \ref{perc}).  

\begin{figure}
\centerline{\mbox{\epsfig{file=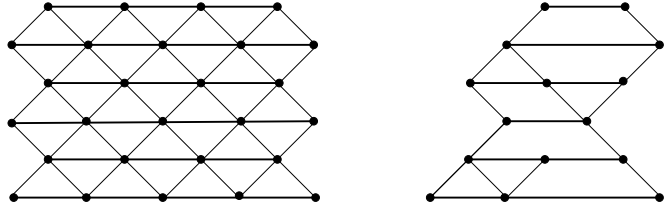}}}
\caption{Maximal and non-maximal evolution of the 1+1 spacetime network of 
\protect\cite{MaSm}. }  
\label{1+1}
\end{figure}

In our context of spin networks, a large number of  arrows on corresponds
to a large number of nodes being spacetime events, which could mean
too much information flowing into a given region, or space expanding
too fast. The interesting situation instead would be when the system
operates close to its $p_c$, where the 1+1 spacetime network of
\cite{MaSm} might look like that on the right of figure \ref{1+1}. In
this case not all nodes are spacetime events at every time step. We
call this case non--maximal, while the situation where all the nodes
are spacetime events will be referred to as maximal.  

\begin{figure}
\centerline{\mbox{\epsfig{file=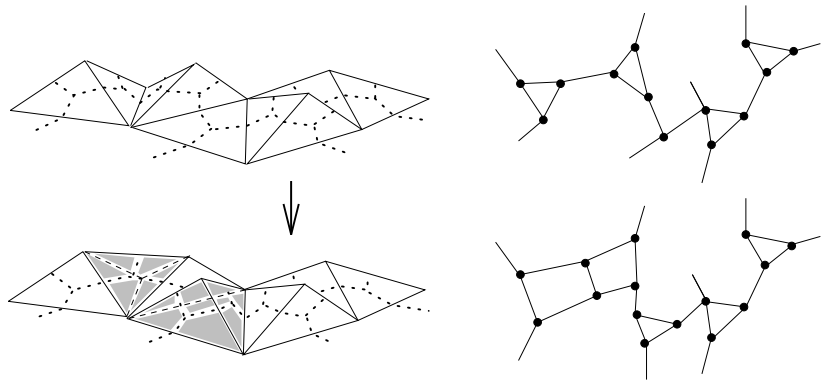}}}
\caption{ Non-maximal 2+1 evolution. The bottom spin network is obtained from the top one by the two shaded spacetime simplices. }
\label{non-max}
\end{figure}

In terms of  triangulations, the non-maximal case is the situation
where, in one time step, spacetime simplices are placed only over {\it
some} spatial sites, thus only changing parts of the spin network,
like multifingered time in the canonical theory. In figure
\ref{non-max} we have placed only two tetrahedra on the given
triangulation of the 2-dimensional space. The result again is that the
growth of the spin network is controlled. This case also slows down
the build-up of spacetime by spacetime simplices, as it takes many
more time steps to achieve a result similar to the maximal case.  
 
\section{Conclusions}

We have illustrated the duality between spin networks and spatial
triangulations, and used this relationship to clarify the recent
causal spin network evolution model, particularly its assumptions of a
connection between valence and dimensionality of space, and its two
evolution rules. We discussed the interpretation of locality and
causality in the triangulated version. This useful duality helped us
to make important extensions to the causal spin network model. In
particular, adding the missing 3--1 move, controlling the growth of
the spin network, allowing the definition of spacelike slices and
varying foliations, and admitting multifingered time. 
The important issue remains the choice of the amplitude functions that
will be consistent with the causality requirement, although the
triangulation formalism provides some indication as to which
symmetries are the desired ones. We may note that we have only worked
with very basic features of the model, for example, we have not
discussed q-deformed spin networks or considered the normalisation of
the amplitudes we propose, which need to be addressed next.  

Since we believe that spin networks in canonical gravity point towards
a discrete spacetime theory, it will be useful to import results and
techniques from simplicial gravity. Clearly, setting up the actual
correspondence between spin networks and triangulations is the first
step. The results of Barett and Foxon in Euclidean and Lorentzian
Regge calculus \cite{BaFo}, together with null-strut calculus
\cite{n-strut}, are likely to shed light on the correct amplitudes for
a Lorentzian spacetime. We shall discuss this elsewhere.  

\begin{figure}
\centerline{\mbox{\epsfig{file=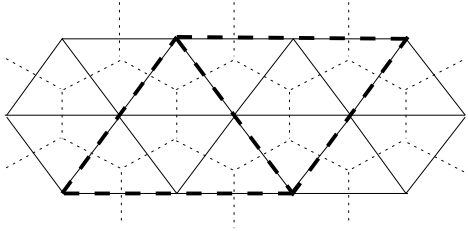}}}
\caption{A grouping of regular trivalent spin network. }
\label{RG}
\end{figure}

Finally, we note that the dual picture makes the increased use of the
graphical representation of the spin network combinatorics more
reliable. For example, in the treatment of spin network evolution as a
critical phenomenon by application of renormalisation group
techniques, the groupings suggested  
by triangulations are more natural and justifiable than when using
spin networks (fig.\ \ref{RG}), although irregularity is still a major
problem in higher dimensions.

\section{Acknowledgments}
I am grateful to John Baez for a great deal of information on
triangulations 
and his explanation of the link to categories. Numerous discussions
with Lee Smolin were essential in clarifying the interpretation
presented here.  Many thanks are also
due to Per Bak, Roumen Borissov, Maya Paczuski, Adam Ritz and Jose
Zapata, and Mike Reisenberger and 
Henri Waelbroeck for their correspondence on spin network causal
evolution. Finally, this paper has benefited from revisions arising
from comments of Louis Crane and Stu Kauffman. 

This work was supported by NSF grant PHY-9514240 to The Pennsylvania State
University and by
the A. S. Onassis foundation. I would also like to thank
Abhay Ashtekar for hospitality at the Center for Gravitational Physics
and Geometry.


\newpage

\begin{thebibliography}{99}

\bibitem{MaSm} F Markopoulou and L Smolin, {\it Causal evolution of spin networks}, gr-qc/9702025.

\bibitem{abhay1}A A Ashtekar, Phys Rev Lett 57 (1986)
2244; Phys Rev D36 (1987) 1587.

\bibitem{tedlee}
T Jacobson and L Smolin, Nucl Phys B 299 (1988).

\bibitem{lp1}C Rovelli and L Smolin, Phys Rev Lett 61 (1988) 1155; Nucl 
Phys B133 (1990) 80.

\bibitem{lp2}R Gambini A Trias, Phys Rev D23 (1981)  553; 
Lett al Nuovo 
Cimento 38 (1983) 497; Phys Rev Lett 53 (1984) 2359; Nucl Phys 
B278 (1986) 436; R Gambini L Leal and  A Trias, Phys Rev D39 
(1989) 3127;  R Gambini, Phys Lett B 255 (1991) 180.  

\bibitem{ls-review}L  Smolin, in {\it ``Quantum Gravity and 
Cosmology''}, eds  J  P\'erez-Mercader {\it et al}, World Scientific, 
Singapore 1992.  

\bibitem{abhay-book} A Ashtekar, {\it ``Non-perturbative canonical 
gravity''}.
 Lecture notes prepared in collaboration with Ranjeet S Tate,  
World Scientific Books, Singapore, 1991.

\bibitem{carlo-review}C Rovelli, Class Quant Grav 
8 (1991) 1613.

\bibitem{sorkin} R Sorkin, {\it Space-time and causal sets}, in Proc
of {\small SILARG VII} Conf, Coyococ, Mexico, Dec 2--7, 1990;
L Bombelli, J Lee, D Meyer and
R D Sorkin,  {\it Spacetime as a causal set} Phys Rev Lett 
59 (1987) 521.

\bibitem{ls}L Smolin, {\it A fixed point for quantum 
gravity},  Nucl Phys B208 (1982) 439; L  Crane and L Smolin, 
{\it Renormalizability of general relativity on a background
of spacetime foam}, Nucl Phys B267 (1986) 714;
{\it Spacetime foam as a universal regulator}, Gen Rel 
Grav 17 (1985) 1209.

\bibitem{penrose} R Penrose, {\it Theory of quantized directions},
unpublished manuscript;  in {\it ``Quantum theory and 
beyond''},   ed T Bastin, Cambridge University Press, 1971.

\bibitem{rosm} C Rovelli and L Smolin, {\it Spin Networks and Quantum
Gravity},   Phys Rev D52 (1995) 5743.  

\bibitem{craneetal} L Crane, L H Kauffman and D N Yetter, {\it State
sum invariants of four manifolds}, hep-th/9409167;  
J C Baez and J Dolan, {\it Higher-dimensional algebra and topological
quantum field theory}, J Math Phys 36, 11 (1995) 6073;
J W Barrett and L Crane, {\it An algebraic interpretation of the
Wheeler-DeWitt equation},  gr-qc/9609030.  

\bibitem{HaPe} B Hasslacher and M J Perry, {\it Spin networks are
simplicial quantum gravity}, Phys Lett 103B:21,1981. 

\bibitem{areavol} C Rovelli and L Smolin, {\it Discreteness of area
and volume in quantum gravity}, Nucl Phys B 442 (1995) 593;
A Ashtekar and J Lewandowski, {\it Quantum Geometry I: area operator},
gr-qc/9602046; 
J Lewandowski, {\it Volume and quantisation}, gr-qc/9602035;
T Thiemann, {\it A length operator in canonical quantum gravity},
gr-qc/9606092. 

\bibitem{jose} J Zapata, {\it A combinatorial approach to
diffeomorphism invariant quantum gauge theories}, gr-qc/9703037; {\it
Combinatorial space from loop quantum gravity}, gr-qc/9703038. 

\bibitem{RR} M Reisenberger and C Rovelli, {\it ``Sum over surfaces''
form of loop quantum gravity},  gr-qc/9612035. 

\bibitem{foam} J Baez, A spin foam model, seminar at Penn State,
February 1997.  

\bibitem{mike} M Reisenberger, {\it A left-handed simplicial action for euclidean general relativity},  gr-qc/9609002.

\bibitem{crane} L Crane, {\it Four-dimensional TQFT: A triptych}, in
Dayton 1992, Proc., Quantum Topology, 116-119; L Crane and I Frenkel,
{\it Four-dimensional topological field theory, Hopf categories and
the canonical bases}, J Math Phys 35 (1994) 5136.

\bibitem{ooguri} H Ooguri, {\it Topological lattice models in
four-dimensions}, Mod Phys Lett A7 (1992) 2799.

\bibitem{mike2} M Reisenberger, private communication.

\bibitem{perc} P Grassberger, Z Phys B 47 (1982) 365; J Stat Phys 79
(1995) 13; H K Janssen, Z Phys B 42 (1981) 151; S Maslov and Y C
Zhang, adapt-org/9601004. 

\bibitem{BaFo} J W Barret and T J Foxon, {\it Semiclassical limits of
simplicial quantum gravity}, Class Quant Grav 11 (1994) 543.

\bibitem{n-strut} A Kheyfets, N J Lafave, W A Miller, {\it } Phys Rev D41 (1990)
3628, 3637.

\end{thebibliography}
\end{document}